\documentclass[12pt,prc,preprint,showpacs,superscriptaddress]{revtex4}
\usepackage{slashed,graphicx}
\tighten
\begin{document}
\title{Nuclear constraints on gravitational waves from rapidly rotating neutron stars}
\author{Aaron Worley}
\altaffiliation[Current address: ]{University of Denver,
Department of Physics and Astronomy, Physics Building, Denver, CO
80208-0183}
\affiliation{Department of Physics, Texas A\&M University-Commerce, P.O. Box 3011,\\
Commerce, TX 75429, U.S.A.}
\author{Plamen G. Krastev}\email[Plamen G. Krastev: ]{pkrastev@sciences.sdsu.edu}
\affiliation{Department of Physics, Texas A\&M University-Commerce, P.O. Box 3011,\\
Commerce, TX 75429, U.S.A.}\affiliation{San Diego State
University, Department of Physics, 5500 Campanile Drive, San Diego
CA 92182-1233}
\author{Bao-An Li}\email[Bao-An Li: ]{Bao-An_Li@tamu-commerce.edu}
\affiliation{Department of Physics, Texas A\&M University-Commerce, P.O. Box 3011,\\
Commerce, TX 75429, U.S.A.}

\date{\today}
\begin{abstract}
Gravitational waves are tiny disturbances in space-time and are a
fundamental, although not yet directly confirmed, prediction of
General Relativity. Rapidly rotating neutron stars are one of the
possible sources of gravitational radiation dependent upon
pulsar's rotational frequency, details of the equation of state of
stellar matter, and distance to detector. Applying an equation of
state with symmetry energy constrained by recent nuclear
laboratory data, we set an upper limit on the strain-amplitude of
gravitational waves emitted by rapidly rotating neutron stars.
\end{abstract}
 \maketitle

\section{Introduction}

General relativity is a theory of gravitation consistent with
special relativity, and in particular with its fundamental
principle that nothing can travel faster than light. This means
that any change in the gravitational field is not felt
instantaneously everywhere in space -- it needs some time to
propagate. In general relativity such disturbances travel at
exactly the same speed as electromagnetic waves in vacuum and are
named {\it gravitational waves}. In this sense gravitational waves
could be regarded as the gravitational equivalent of
electromagnetic radiation -- just as electromagnetic waves carry
information about rearrangement of electric charges and currents,
gravitational waves carry similar information about rearrangement
of masses in space.

Although a fundamental prediction of general relativity,
gravitational waves are yet to be detected directly. Such an
observation would have profound consequences for our basic
understanding of matter, space and time, and how they couple to
each other. Because gravity interacts extremely weakly with
matter, gravitational waves would provide detailed information
about their sources presently hidden, or dark, to current
electromagnetic observations~\cite{Maggiore:2007,Flanagan:2005yc}.
Due to their potential as a unique probe of new physics and their
fundamental nature, the search for gravitational waves has
attracted enormous effort over the last few years by the
LIGO~\cite{Abbott:2004ig}, VIRGO (e.g.,
Ref.~\cite{Acernese:2007zzb}), and GEO (e.g.,
Ref.~\cite{Abbott:2004NIMPR}) collaborations. Moreover, LISA (the
Laser Interferometric Space Antenna) is currently being jointly
designed by NASA in the United States and ESA (the European Space
Agency), and is scheduled to be launched into orbit around 2018
providing an unprecedented instrument for gravitational waves
search and detection~\cite{Flanagan:2005yc}.

Rapidly rotating neutron stars could be one of the major
candidates for sources of continuous gravitational radiation in
the frequency bandwidth of ground-based laser interferometric
detectors such as LIGO and VIRGO. To emit gravitational waves over
extended period of time, a rotating object bound by gravity must
have some kind of long-living axial
asymmetry~\cite{Jaranowski:1998qm}. In the literature several
possible mechanism leading to such deformations have been proposed
(e.g., Ref.\cite{Jaranowski:1998qm}). Among the factors
contributing to axial asymmetries, relevant to our present study,
is the anisotropic stress which may accumulate during the
crystallization period of the neutron star crust and thus support
axial distortions~\cite{PPS:1976ApJ}. On the other hand,
gravitational wave strain amplitude depends on the degree to which
the neutron star is deformed from axial asymmetry, which, in turn,
is dependent upon the details of the equation of state (EOS) of
neutron-rich stellar matter. Presently, the EOS of dense,
neutron-rich matter is still rather uncertain mainly due to the
poorly known density dependence of the nuclear symmetry energy
$E_{sym}(\rho)$, e.g.~\cite{Lattimer:2004pg}. Heavy-ion reactions
with radioactive beams could provide unique means to constrain the
uncertain density behavior of the nuclear symmetry energy and thus
the EOS of neutron-rich nuclear matter,
e.g.~\cite{Li:1997px,Li:1997rc,Li:2000bj,Li:2002qx,LCK08}.
Although there are important differences between dense,
neutron-rich matter produced in the laboratory and that found in
the interior of neutron stars, it is still very useful and
insightful to study the astrophysical implications of the
constrained EOS -- in particular, its impact on the neutron star
properties and the resultant gravitational waves. Following a
recent work~\cite{Krastev:2008PLB} in which we have constrained
the gravitational waves expected from elliptically deformed {\it
slowly rotating} pulsars, in this paper we extend our studies to
{\it rapidly rotating} neutron stars. Applying several nucleonic
EOSs, we calculate the strain amplitude of gravitational waves
expected from the fastest neutron stars known as of today.
Particular attention is paid to predictions with an EOS with
symmetry energy constrained by very recent nuclear laboratory
data. These results set an upper limit on the strain amplitude of
gravitational radiation expected from rapidly rotating neutron
stars. This paper is organized in the following manner. After the
introductory remarks in this section, in the next section we
discuss briefly the formalism for calculating the gravitational
wave strain amplitude for rapidly rotating neutron stars. Our
results are presented and discussed in section 3. We conclude in
section 4 with a short summary.

\section{Formalism}

In what follows we review briefly the formalism used to calculate
the gravitational wave strain amplitude. Here we are specifically
concerned with gravitational waves (GWs) from rapidly rotating
neutron stars. (Details about computing the strain amplitude in
the case of slowly rotating pulsars can be found in
Ref.~\cite{Krastev:2008PLB} and references therein.) A spinning
neutron star is expected to emit GWs if it is not perfectly
symmetric about its rotational axis. As already mentioned,
non-axial asymmetries may be produced through several mechanisms
such as elastic deformations of the solid crust or core or
distortion of the whole star by extremely strong misaligned
magnetic fields~\cite{BG:1996AA}. Such processes generally result
in a triaxial neutron star configuration~\cite{Abbott:2004ig}
which, in the quadrupole approximation and with rotation and
angular momentum axes aligned, would cause gravitational waves at
{\it twice} the star's rotational frequency~\cite{Abbott:2004ig}.
These waves have characteristic strain amplitude at the Earth's
vicinity (assuming an optimal orientation of the rotation axis
with respect to the observer) of~\cite{HAJS:2007PRL}
\begin{equation}\label{Eq.1}
h_0=\frac{16\pi^2G}{c^4}\frac{\epsilon I_{zz}\nu^2}{r},
\end{equation}
where $\nu$ is the neutron star rotational frequency, $I_{zz}$ its
principal moment of inertia, $\epsilon=(I_{xx}-I_{yy})/I_{zz} $
its equatorial ellipticity, and $r$ its distance to Earth.

Provided the spin-down rate, $\dot{\nu}$, for a given pulsar is
known, $h_0$ could be written as \cite{Abbott:2007PRD}
\begin{equation}\label{Eq.2}
h_0^{sd}=\frac{5}{2}\left(\frac{GI_{zz}|\dot{\nu}|}{c^3r^2\nu}\right)^{1/2},
\end{equation}
which provides an alternative way for calculating the
gravitational wave strain amplitude. Here we should mention that
in such calculations one assumes that the {\it only} mechanism
contributing to the pulsar's observed spin-down is gravitational
radiation. However, other mechanisms could also account for the
star's observed decrease in rotational frequency such as magnetic
dipole radiation, and particle acceleration in the
magnetosphere~\cite{Abbott:2008APL}. Despite these uncertainties,
calculations of gravitational wave strain amplitude through
Eq.~(\ref{Eq.2}) are still important because, in addition to
providing a rather conservative upper limit on the expected
gravitational radiation, they also serve to estimate another very
uncertain but important quantity -- the ellipticity $\epsilon$,
which is a measure of the neutron star deformation. ($\epsilon$
could be evaluated through combining Eqs.~(\ref{Eq.1}) and
(\ref{Eq.2}).) Provided one knows the exact rotational frequency,
principal moment of inertia, distance to detector, and ellipticity
(or spin-down rate), Eq.~(\ref{Eq.1}) (or Eq.~(\ref{Eq.2})) can be
used to calculate the gravitational wave strain amplitude. These
estimates are then to be compared with the current upper limits
for sensitivity of the laser interferometric observatories (e.g.,
LIGO). In the present work we use the RNS
code~\cite{Stergioulas:1994ea}, written and made available to the
public by Nikolaos Stergioulas, to calculate the principal moment
of inertia (and other properties) of rapidly rotating neutron
stars. (The $RNS$ code is available as a public domain program at
http://www.gravity.phys.uwm.edu/rns/.)

Here we should point out that the fastest pulsars presently known
(PSR B1937+21, PSR J1748-2446ad, and XTE J1739-285) have
rotational frequencies in the upper end of the current detection
limit of LIGO ($\sim 200Hz$). On the other hand, Eq.~(\ref{Eq.1})
implies that {\it rapidly} rotating neutron stars emit stronger
gravitational waves, and therefore it is important to study the
gravitational waves expected from such neutron stars.

\section{Results and discussion}

\begin{figure}[!b]
\centering
\includegraphics[totalheight=3.1in]{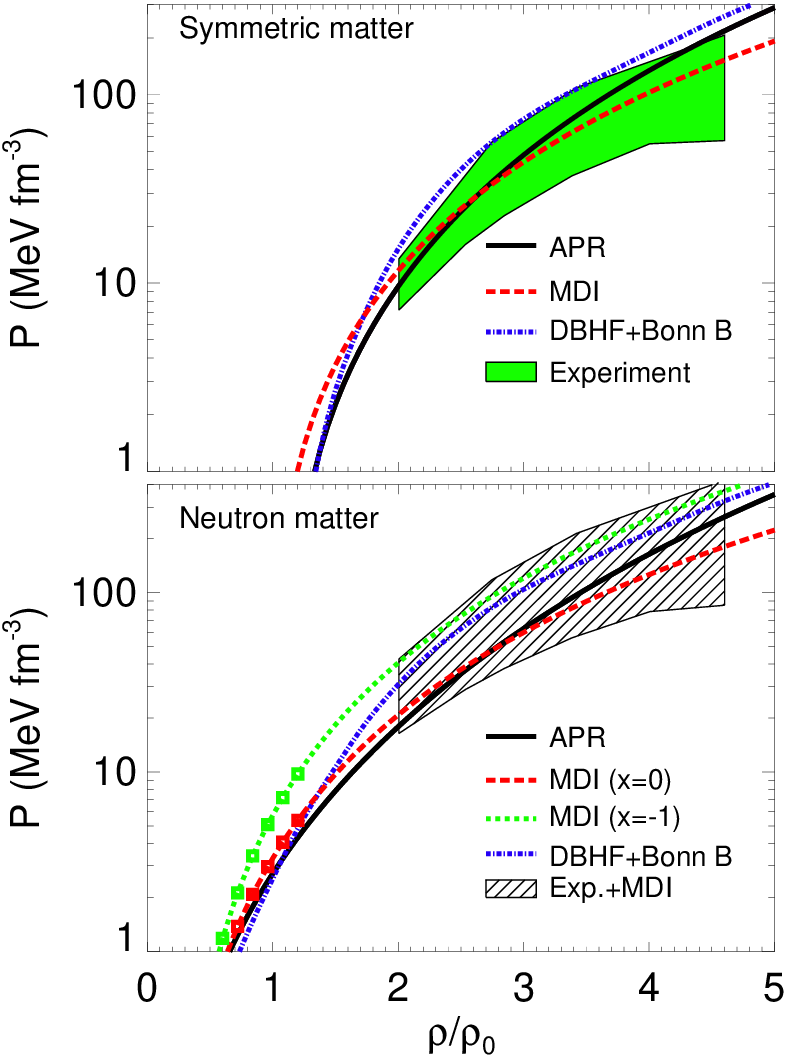}
\vspace{5mm} \caption{(Color online) Pressure as a function of
density for symmetric (upper panel) and pure neutron (lower panel)
matter. Taken from Ref. \cite{Krastev:2008PLB}. (Details about the
figure can be found in Ref.~\cite{Krastev:2008PLB}.)}\label{Fig.1}
\end{figure}

\begin{table}[!b]
\caption{Properties of the {\it rapidly rotating} pulsars
considered in this study. The first column identifies the pulsar.
The remaining columns exhibit the following quantities: rotational
frequency; first derivative of the rotational frequency; distance
to Earth; corresponding reference.} \vspace{3mm} \centering
\begin{tabular}{lcccc}
Pulsar        & $\nu(Hz)$ & $\dot{\nu}(Hz$ $s^{-1})$ & $r(kpc)$ &Reference  \\
\hline\hline
PSR B1937+21      & 641.93   & $-$4.33$\times$ $10^{-14}$  &  3.60  &  \cite{Backer:1982,Cusmano:2004NPPS}     \\
PSR J1748-2446ad  & 716.35   & --                          &  8.70  &  \cite{Hessels:2006ze,Manchester:2005AJ} \\
XTE J1739-285     & 1122.00  & --                          & 10.60  &  \cite{Kaaret:2006gr}                    \\
\hline
\end{tabular}
\end{table}

In this work, we calculate the gravitational wave strain amplitude
$h_0$ for the fastest pulsars currently known employing several
nucleonic equations of state. We assume a simple model of stellar
matter of nucleons and light leptons (electrons and muons) in
beta-equilibrium. Details about the constrained equation of state
calculated with the Momentum Dependent Interaction
(MDI)~\cite{Das:2002fr} applied in the present paper can be found,
for instance, in Refs.~\cite{LCK08,Krastev:2008PLB}. Here we
recall briefly its main features. For many astrophysical studies
(as those in this paper), it is more convenient to express the EOS
in terms of the pressure as a function of density and isospin
asymmetry. In Fig.~\ref{Fig.1} we show pressure as a function of
density for symmetric (upper panel) and pure neutron matter (lower
panel). The EOS of symmetric nuclear matter with the MDI
interaction is constrained by the available data on collective
flow in relativistic heavy-ion collisions. The parameter $x$ is
introduced in the single-particle potential of the MDI EOS to
account for the largely uncertain density dependence of the
nuclear symmetry energy $E_{sym}(\rho)$ as predicted by various
many-body frameworks and models of the nuclear force. Since it was
demonstrated by Li and Chen~\cite{Li:2005jy} and Li and
Steiner~\cite{Li:2005sr} that only equations of state with $x$ in
the range between -1 and 0 have symmetry energy consistent with
the isospin-diffusion laboratory data and measurements of the skin
thickness of $^{208}Pb$, we therefore consider only these two
limiting cases in calculating boundaries of the possible
(rotating) neutron star configurations. It is also important to
mention that the symmetry energy extracted very recently from
isoscaling analysis of heavy-ion reactions is consistent with the
MDI calculation of the EOS with $x=0$~\cite{Shetty:2007}. The MDI
EOS has been applied to constrain the neutron star
radius~\cite{Li:2005sr} with a suggested range compatible with the
best estimates from observations. It has been also used to
constrain a possible time variation of the gravitational constant
$G$ \cite{Krastev:2007en} via the {\it gravitochemical heating}
approach developed by Jofre et al.~\cite{Jofre:2006ug}. More
recently we applied the MDI EOS to constrain the global properties
of (rapidly) rotating neutron stars~\cite{WKL:2008ApJ,KLW2} and
the strain amplitude of the gravitational waves expected from
elliptically deformed {\it slowly rotating}
pulsars~\cite{Krastev:2008PLB}. In addition to the MDI EOS, in
Fig.~\ref{Fig.1} we show results by Akmal et
al.~\cite{Akmal:1998cf} with the $A18+\delta\upsilon+UIX*$
interaction (APR) and very recent Dirac-Brueckner-Hartree-Fock
(DBHF) calculations~\cite{Sammarruca:2008iu} with Bonn B
One-Boson-Exchange (OBE) potential (DBHF+Bonn
B)~\cite{Machleidt:1989}. (Older calculations of the DBHF+Bonn B
EOS can be found in~\cite{Alonso:2003aq,Krastev:2006ii}.) Below
the baryon density of approximately $0.07fm^{-3}$ the equations of
state applied here are supplemented by a crustal EOS, which is
more suitable for the low density regime. Namely, we apply the EOS
by Pethick et al.~\cite{PRL1995} for the inner crust and the one
by Haensel and Pichon~\cite{HP1994} for the outer crust. At the
highest densities we assume a continuous functional for the EOSs
employed in this work.

\begin{figure}[!t]
\centering
\includegraphics[totalheight=4.0in]{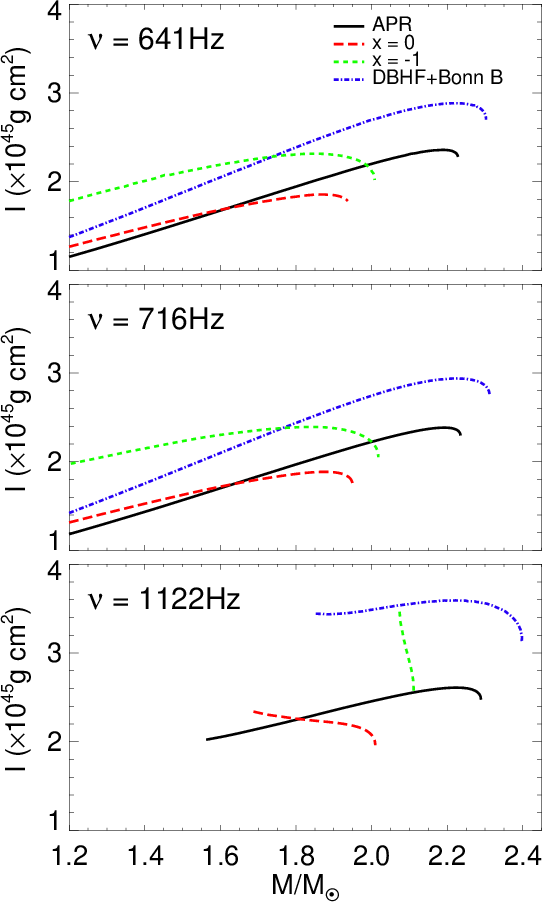}
\vspace{5mm} \caption{(Color online) Neutron star moment of
inertia as a function of stellar mass. Results are partially taken
from Ref.~\cite{WKL:2008ApJ}.}\label{Fig.2}
\end{figure}

In this paper we study gravitational waves emitted from rapidly
rotating neutron stars. Specifically, we examine configurations
rotating at 641, 716, and 1122 Hz. These frequencies represent the
three fastest pulsars discovered as of today. The properties of
these neutron stars (of interest in this study) are summarized in
Table 1. In Fig.~\ref{Fig.2} we show the neutron star moment of
inertia as a function of stellar mass. The upper panel shows the
moment of inertia of PSR B1937+21, while the middle and lower
panels display the moments of inertia of PSR J1748-2446ad and XTE
J1739-285 respectively. As already observed previously, the moment
of inertia increases with rotational frequency, while the range of
possible neutron star configurations decreases (see, for instance,
Refs.~\cite{KLW2,WKL:2008ApJ}).

\subsection{Rotations at 641 Hz}

\begin{figure}[t!]
\centering
\includegraphics[height=5.5cm]{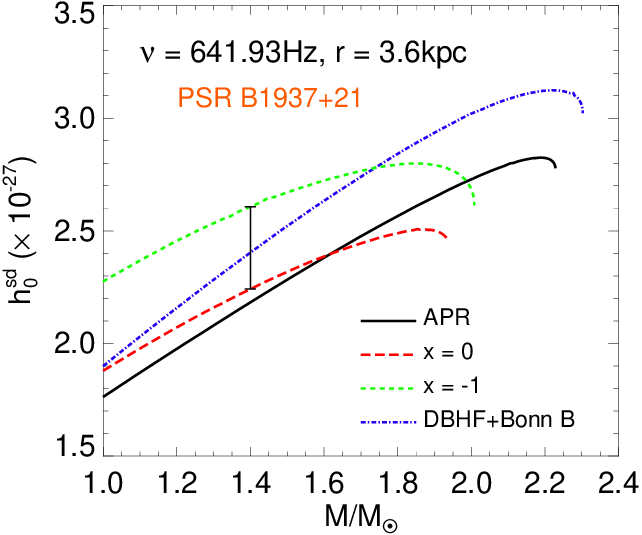}
\includegraphics[height=5.5cm]{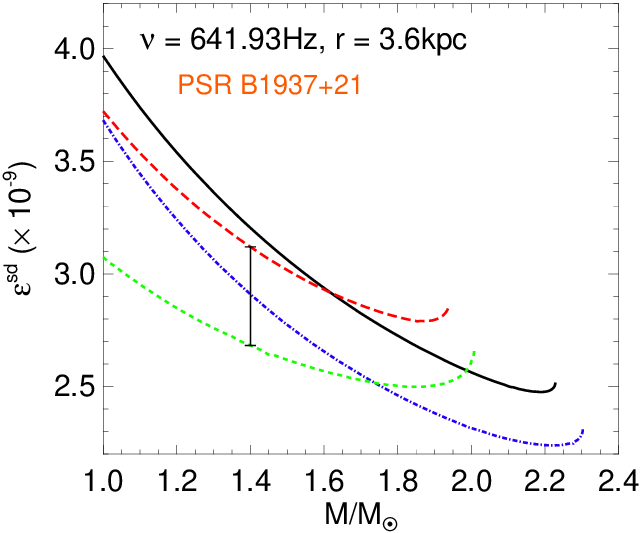}
\caption{(Color online) Neutron star gravitational wave strain
amplitude (left panel) and ellipticity (right panel) as a function
of stellar mass. For computing $h_0^{sd}$ Eq.~(\ref{Eq.2}) has
been used. The ellipticity, $\epsilon^{sd}$, has been calculated
by combining Eqs. (\ref{Eq.1}) and (\ref{Eq.2}). The error bars
between the $x=0$ and $x=-1$ EOSs, in both frames, provide a limit
on the strain amplitude of the gravitational waves to be expected
from this neutron star and its ellipticity, and show a specific
case for stellar models of $1.4M_{\odot}$.} \label{Fig.2}
\end{figure}

In this subsection we study gravitational waves from neutron star
models rotating a 641 Hz~\cite{Backer:1982}, which was the
rotational frequency of the fastest pulsar (PSR B1937+21) for 23
years before the discovery by Hessels et al.~\cite{Hessels:2006ze}
in 2006. Since its first observation in 1982, this pulsar has been
studied extensively and an observed spin-down rate has been
measured (see Table 1). Using the spin-down rate, we placed an
upper limit on the gravitational wave strain amplitude from PSR
B1937+21, assuming the only mechanism contributing to the observed
spin-down is gravitational radiation. This is a rather simplistic
approach because, as already mentioned, there are other mechanisms
that could/would account for the star's observed decrease in
rotational frequency. For instance, young pulsars could exhibit
significant amount of energy loss through electromagnetic
processes. Nevertheless, this method still provides a conservative
upper bound to the amplitude of the GWs that can be emitted. The
spin-down rate corresponds to a loss in kinetic energy at a rate
of $\dot{E}$ = $4\pi^2I_{zz}\nu|\dot{\nu}|\sim$[0.6 -- 3.1]
$\times$ 10$^{36}erg/s$, depending on the EOS. Assuming that the
full amount of this energy loss is being radiated away is in the
form of gravitational radiation, the gravitational wave strain
amplitude can be calculated through Eq.~(\ref{Eq.2}). Similar
calculations for this pulsar and others with an observed spin-down
rate have been done in the
past~\cite{Abbott:2004ig,Abbott:2007PRD}. However, such
calculations simply used the ``fiducial''  value of $10^{45}$$g$
$cm^2$ for the moment of inertia $I_{zz}$, while here we calculate
the moment of inertia of PSR B1937+21 numerically with the RNS
code for each EOS (upper panel of Fig.~2).

The gravitational wave strain amplitude of PSR B1937+21 is shown
in the left panel of Fig.~3. Because the MDI EOS is constrained by
available nuclear laboratory data, our results with the $x=0$ and
$x=-1$ EOSs allowed us to place a rather conservative {\it upper}
limit on the gravitational waves to be expected from this pulsar,
provided the {\it only} mechanism accounting for its spin-down
rate is gravitational radiation. Under these circumstances, the
upper limit of the strain amplitude, $h_0^{sd}$, for neutron star
models of $1.4M_{\odot}$ is in the range
$h_0^{sd}=[2.24-2.61]\times 10^{-27}$. Similarly, we have
constrained the upper limit of the ellipticity of PSR B1937+21 to
be in the range $\epsilon^{sd}=[2.68-3.12]\times 10^{-9}$ (Fig.~3,
right panel).

\begin{figure}[!t]
\centering
\includegraphics[totalheight=6cm]{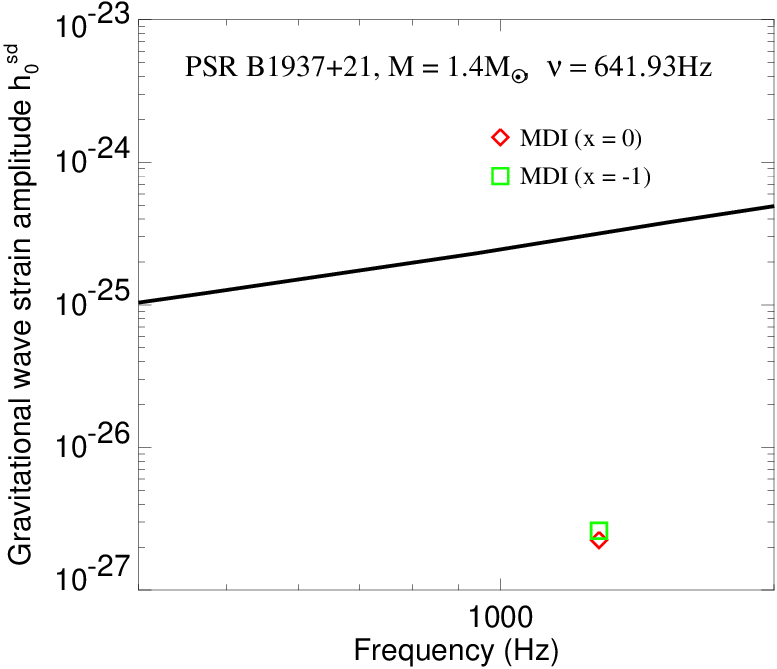}
\vspace{5mm} \caption{(Color online) Gravitational wave strain
amplitude as a function of the gravitational wave frequency. The
characters denote the strain amplitude of the GWs expected to be
emitted from neutron stars spinning at $641.93Hz$ with mass
$1.4M_{\odot}$. Solid line (adapted from
Ref.~\cite{Abbott:2004ig}) denotes the current upper limit of the
LIGO sensitivity.}
\end{figure}

In Fig.~4 we take another view of the results shown in the left
frame of Fig.~3. We display the maximal GW strain amplitude as a
function of the GW frequency and compare our predictions with the
best current detection limit of LIGO. The specific case shown is
for neutron star models with mass $1.4M_{\odot}$ computed with the
$x=0$ and $x=-1$ EOSs. Since these EOSs are constrained by the
available nuclear laboratory data they provide a limit on the
possible neutron star configurations and thus gravitational
emission from them. From Fig.~4 it could concluded that the GWs
emitted by PSR B1937+21 associated with the pulsar's spin-down are
well below the current detection limit of LIGO. On the other hand,
Eq.~(\ref{Eq.2}), through which the GW strain amplitude has been
computed, does not take into account other processes contributing
to the decrease in angular velocity of the star. Moreover, Haskell
et al.~\cite{HJA:2006MNRAS} has shown that the ellipticity of an
isolated neutron star, which is a measure of stellar deformation
from axial symmetry, could be as large as $\sim 10^{-6}$. This
value of $\epsilon$ is 3 orders of magnitude larger than the
estimates from the pulsar's spin-down shown in Fig.~3 (right
panel), and would bring the GW strain amplitude within the current
detection range of the interferometric detectors (e.g., LIGO). The
large uncertainty of the ellipticity is dictated by the very
poorly known value of the breaking strain of the neutron star
crust $\bar{\sigma}_{br}$. It falls in the very wide range
$\bar{\sigma}_{br}=[10^{-5}-10^{-2}]$ and has to be pinned down
from more realistic calculations. However, this is not an easy
task as it requires detailed microscopic description of matter in
the neutron star crust (see, e.g. Ref.~\cite{Horowitz:2008vf}).

\subsection{Rotations at 716 and 1122 Hz}

\begin{figure}[b!]
\centering
\includegraphics[height=5.5cm]{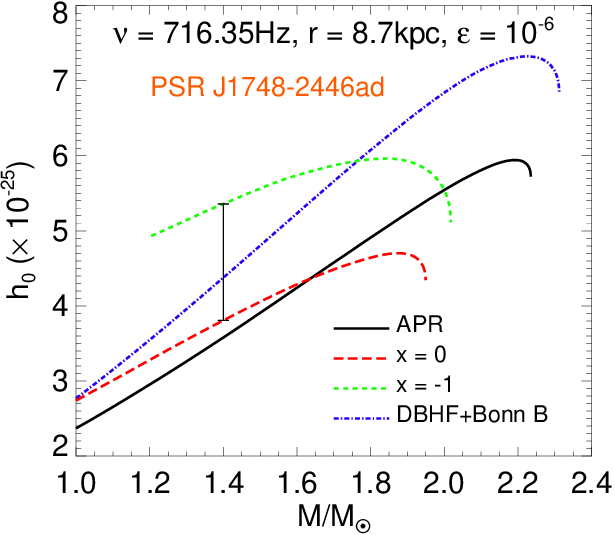}
\includegraphics[height=5.5cm]{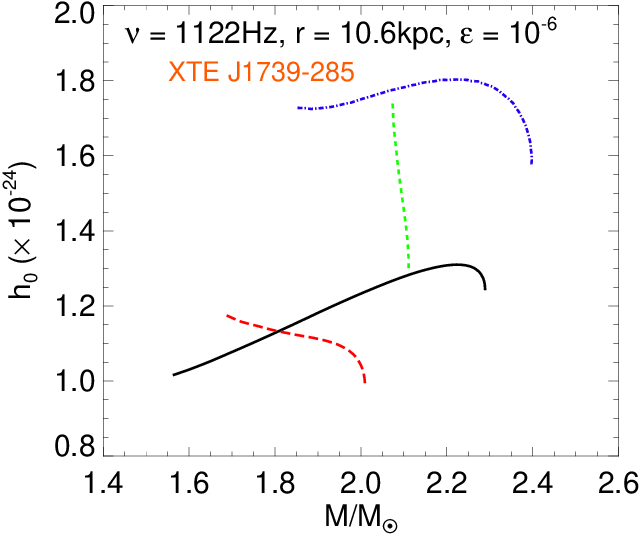}
\caption{(Color online) Gravitational wave strain amplitude,
$h_0$, versus neutron star mass for models rotating at 716 Hz
(left frame) and 1122 Hz (right frame). The error bar in the left
frame between the $x=0$ and $x=-1$ EOSs provide a limit on the
strain amplitude of the GWs expected from PSR J1748-2446ad and
show a specific case for stellar models of $1.4M_{\odot}$.}
\end{figure}

In this subsection we examine the gravitational waves emitted from
neutron star models rotating at 716 and 1122 Hz which are the
rotational frequencies of PSR J1748-2446ad and XTE J1739-285
respectively. Presently the spin-down rates of these objects are
not available and therefore estimates of GW strain amplitude, and
in turn ellipticity, through Eq.~(\ref{Eq.2}) are not possible. We
calculate $h_0$ through Eq.~(\ref{Eq.1}) with $I_{zz}$ computed
with the RNS code and $\epsilon=10^{-6}$. The specific value of
$\epsilon$ chosen here maximizes $h_0$ and is consistent with the
{\it largest} ``mountain'' expected on a isolated neutron
star~\cite{HJA:2006MNRAS}. We show the GW strain amplitude as a
function of stellar mass in Fig.~5. Because ellipticity is fixed
in the present calculation, $h_0$ follows qualitatively the mass
dependence of the moment of inertia (Fig.~2). Our results with the
MDI EOS (x = 0, -1) set a constraint on the {\it upper} limit for
$h_0$ of the GWs expected to be emitted from PSR J1748-2446ad. If
the pulsar's mass is $\sim 1.4M_{\odot}$ then
$h_0=[3.81-5.36]\times 10^{-25}$. Similarly, we have placed a
limit on the strain amplitude of the GWs generated by XTE
J1739-285. As previously shown~\cite{KLW2}, only narrow range of
neutron star models is possible at such large frequencies (with
details depending on the specific EOS of stellar matter).
Moreover, the mass of XTE J1739-285 has been shown~\cite{KLW2} to
exceed the canonical value of $1.4M_{\odot}$. From Fig.~5 (right
frame) we conclude that the {\it absolute upper} bound of $h_0$
for this neutron star is somewhere in the range
$h_0=[1.17-1.74]\times 10^{-24}$, if the pulsar's mass falls in
the range $M=[1.69-2.07]M_{\odot}$.

These estimates do not take into account the uncertainties in the
distance measurements as well as the uncertainties in the
orientation of the star with respect to Earth, and they are also
regarded as upper limits as a result of the ellipticity chosen as
a maximum. The results shown in Fig.~5 also reveal that $h_0$
depends on the EOS. However, the exact dependence could be only
established, if the stellar ellipticity is also calculated
consistently. Such calculations would require an exact calculation
of the quadrupole moment of rapidly rotating neutron stars which
is not trivial (see, e.g. Ref.~\cite{Laarakkers:1999ApJ}).
Laarakkers and Poisson~\cite{Laarakkers:1999ApJ} have extended the
RNS code to calculate also the quadrupole moment, $\Phi_{22}$. The
ellipticity then could be calculated as
$\epsilon=(8\pi/15)^{1/2}\Phi_{22}/I_{zz}$. This would yield,
however, unrealistically large values for $\epsilon$ and, in turn,
$h_0$ (as verified in preliminary calculations).

\begin{figure}[t!] \centering
\includegraphics[height=5.5cm]{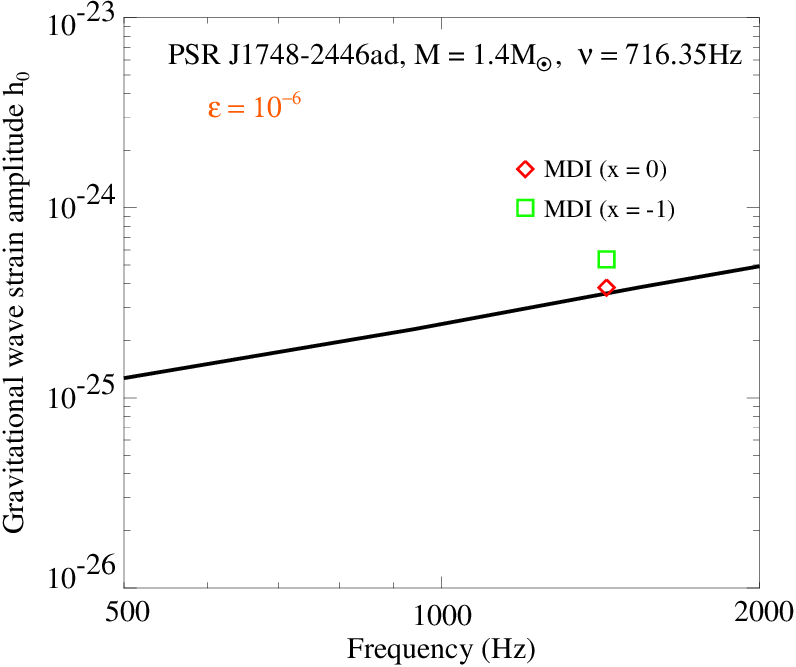}
\includegraphics[height=5.5cm]{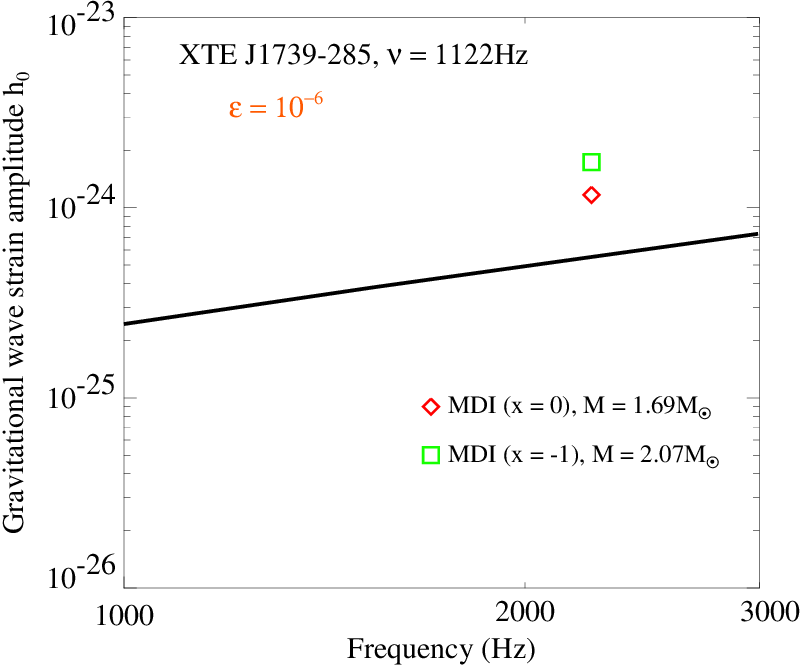}
\caption{(Color online) Gravitational wave strain amplitude as a
function of the gravitational wave frequency for neutron star
models spinning at 716 Hz (left panel) and 1122Hz (right panel).
Conventions and labelling as in Fig.~4.}
\end{figure}

At the end, in Fig.~6 we examine the results shown in Fig.~5 from
a different perspective. We display the maximum GW strain
amplitude as a function of GW frequency ($f_{gw}=2\nu$) and
compare our predictions with the best current detection limit of
LIGO. As already mentioned, the ellipticity has been chosen such
that to maximize $h_0$. For neutron star models rotating at 716 Hz
(left frame) we show specific case of stellar configurations of
$1.4M_{\odot}$. On the other hand, in the case of rapid rotation
at 1122 Hz, $1.4M_{\odot}$ configurations do not exist as stable
models should be at least $\sim1.7M_{\odot}$ (for the $x=0$ EOS).
Therefore, in Fig.~6 (right frame) we show $h_0$ for the {\it
lowest} mass neutron star models. Moreover, as recently
discussed~\cite{Krastev:2008PLB} low-mass neutron stars are
expected to generate stronger gravitational waves (at the same
rotational frequency). This is because such models are less
compact (and gravitationally bound) which could result in a
greater susceptibility of deformation by various mechanisms and
phenomena, and ultimately lead to a stronger GW strain amplitude.
The results shown in Fig.~6 would suggest that presently the
gravitational radiation from these pulsars should be within the
detection capabilities of LIGO. Several factors could contribute
to the fact that, on contrarily, such a detection has not been
made yet. First, for calculating $h_0$ shown in Fig.~6 we have
used the maximum ellipticity $\epsilon=10^{-6}$. On the other
hand, $\epsilon$ could be significantly lower~\cite{HJA:2006MNRAS}
which would bring $h_0$ to lower values and out of the LIGO's
current detection range (see also the previous subsection).
Second, in the present study we assume a very simple model of
stellar matter consisting only beta equilibrated nucleons and
light leptons (electrons and muons). On the other hand, in the
core of neutron stars conditions are such that other more exotic
species of particles could readily abound. Such novel phases of
matter would soften considerably the EOS of stellar medium
\cite{BBS:2000PRC} leading to ultimately more compact and
gravitationally tightly bound objects which could withstand larger
deformation forces (and torques). Lastly, the existence of quark
stars, truly exotic self-bound compact objects, is not excluded
from further considerations and studies. Such stars would be able
to resist huge forces (such as those resulting from extremely
rapid rotation beyond the Kepler, or mass-shedding, frequency) and
as a result retain their axial symmetric shapes effectively
dumping the gravitational radiation (e.g. \cite{Weber:1999a}). We
leave this discussion by simply recalling that Eq.~(\ref{Eq.1})
implies that the best possible candidates for gravitational
radiation (from spinning relativistic stars) are rapidly rotating
pulsars relatively close to Earth ($h_0\sim \nu^2/r$), thus,
further understanding of the neutron star's ellipticity under such
rotational conditions are important for more realistic
calculations.

\section{Summary}

In this paper we have reported predictions on the upper limit of
the strain amplitude of the gravitational waves expected to be
emitted from the fastest pulsars presently known. By applying an
EOS with symmetry energy constrained by recent nuclear laboratory
data, we obtained an upper limit on the gravitational - wave
signal to be expected from PSR B1937+21, PSR J1748-2446ad, and XTE
J1739-285. These predictions serve as the first direct nuclear
constraint on the gravitational waves from {\it rapidly} rotating
neutron stars.

\section*{Acknowledgements}

The authors gratefully acknowledge support from the National
Science Foundation under Grants No. PHY0652548 and No. PHY0757839,
the Research Corporation under Award No. 7123 and the Texas
Coordinating Board of Higher Education Grant No. 003565-0004-2007.
We thank F. Sammarruca for providing the updated DBHF + Bonn B
EOS.


\begin{thebibliography}{100}

\bibitem{Maggiore:2007}M.~Maggiore, Nature {\bf 447},
651 (2007).

\bibitem{Flanagan:2005yc}
  E.~E.~Flanagan and S.~A.~Hughes,
  New J.\ Phys.\  {\bf 7}, 204 (2005).

\bibitem{Abbott:2004ig}
  B.~Abbott {\it et al.}  [LIGO Scientific Collaboration],
  Phys.\ Rev.\ Lett.\  {\bf 94}, 181103 (2005); Phys.\ Rev.\  D {\bf
  76}, 042001 (2007).

\bibitem{Acernese:2007zzb}
  F.~Acernese {\it et al.},
  Class.\ Quant.\ Grav.\  {\bf 24}, S491 (2007).

\bibitem{Abbott:2004NIMPR}B.~Abbott et al. (LSC), Nucl. Instrum. Meth. Phys. Res. A {\bf
517}, 154-179 (2004).

\bibitem{Jaranowski:1998qm}
  P.~Jaranowski, A.~Krolak and B.~F.~Schutz,
  Phys.\ Rev.\  D {\bf 58}, 063001 (1998).

\bibitem{PPS:1976ApJ}V.~R.~Padharipande, D.~Pines, and
R.~A.~Smith, Astrophys. J. {\bf 208}, 550--566 (1976).

\bibitem{Krastev:2007en}P.~G.~Krastev, B.-A. Li, Phys. Rev. C {\bf 76}, 055804 (2007).

\bibitem{KLW2}P.~G.~Krastev, B.-A. Li, and A.~Worley, Astrophys. J. {\bf 676}, 1170–1177 (2008).

\bibitem{WKL:2008ApJ}A. Worley, P. G. Krastev, and B.-A. Li,
Astrophys. J. {bf 685}, 390 (2008).

\bibitem{Krastev:2008PLB}Plamen G. Krastev, Bao-An Li, and Aaron
Worley, Phys. Lett. B {\bf 668}, 1 (2008).

\bibitem{Lattimer:2004pg}J.~M.~Lattimer and M.~Prakash, Science {\bf 304}, 536 (2004).

\bibitem{Li:1997px}B.-A. Li, C.~M.~Ko, and W.~Bauer, Int. J. Mod. Phys.
E7, 147 (1998).

\bibitem{Li:1997rc}B.-A. Li, C.~M.~Ko, and Z.-Z.~Ren, Phys. Rev. Lett. {\bf 78}, 1644 (1997).

\bibitem{Li:2000bj}B.-A. Li, Phys. Rev. Lett. {\bf 85}, 4221 (2000).

\bibitem{Li:2002qx}B.-A. Li, Phys. Rev. Lett.  {\bf 88}, 192701 (2002).

\bibitem{LCK08}B.-A. Li, L.-W. Chen and C.M. Ko, Phys. Rep. {\bf 464}, 113 (2008).

\bibitem{BG:1996AA}S.~Bonazzola and E.~Gourgoulhon, Astron. Astrophys. {\bf 312}, 675 (1996).

\bibitem{HAJS:2007PRL}B. Haskell, N. Andersson, D. I. Jones, and L.
Samuelsson, Phys. Rev. Lett. {\bf 99}, 231101 (2007).

\bibitem{Abbott:2007PRD}B.~Abbott et al., Phys. Rev. D {\bf 76}, 042001 (2007).

\bibitem{Abbott:2008APL}B.~Abbott et al., Astrophys. J. Lett. (2008), in press;
arXiv:0805.4758.

\bibitem{Das:2002fr}C.~B.~Das, S.~D.~Gupta, C.~Gale, and B.-A. Li, Phys. Rev. C {\bf 67}, 034611 (2003).

\bibitem{Li:2005jy}B.-A. Li and L.-W. Chen, Phys. Rev. C {\bf 72}, 064611 (2005).

\bibitem{Li:2005sr}B.-A. Li and A. W. Steiner, Phys. Lett. B {\bf 642}, 436 (2006).

\bibitem{Shetty:2007}D.~Shetty, S. J. Yennello, and G. A. Souliotis, Phys. Rev. C {\bf 75}, 034602 (2007).

\bibitem{Jofre:2006ug}P.~Jofre, A.~Reisenegger, and R.~Fernandez, Phys. Rev. Lett. {\bf 97}, 131102 (2006).

\bibitem{Akmal:1998cf}A.~Akmal, V.~R.~Pandharipande, D.~G.~Ravenhall, Phys. Rev. C {\bf 58}, 1804 (1998).

\bibitem{Sammarruca:2008iu}
  F.~Sammarruca and P.~Liu,
  arXiv:0806.1936 [nucl-th].

\bibitem{Alonso:2003aq}D.~Alonso and F.~Sammarruca, Phys. Rev. C {\bf 67}, 054301 (2003).

\bibitem{Krastev:2006ii}P.~G.~Krastev and F.~Sammarruca, Phys. Rev. C {\bf 74}, 025808 (2006).

\bibitem{Machleidt:1989}R.~Machleidt, Adv. Nucl. Phys. {\bf 19}, 189 (1989).

\bibitem{PRL1995}C.~J.~Pethick, D.~G.~Ravenhall, and C.~P.~Lorenz, Nucl. Phys. A {\bf 584}, 675 (1995).

\bibitem{HP1994}P.~Haensel and B.~Pichon, Astron. Astrophys. {\bf 283}, 313 (1994).

\bibitem{FPI:1984Nature}J.~L.~Friedman, L.~Parker, and J.~R~Ipser, Nature {\bf 312}, 25 (1984).

\bibitem{HJA:2006MNRAS}B.~Haskell, D.~I.~Jones, and N.~Andersson, Mon. Not. R. Astron. Soc. {\bf 373}, 1423 (2006).

\bibitem{BBS:2000PRC}M.~Baldo, G.~F.~Burgio, and  H.-J.~Schulze, Phys Rev. C {\bf 61}, 055801 (2000).

\bibitem{Weber:1999a}F.~Weber, Pulsars as Astrophysical Laboratories for Nuclear and
Particle Physics (1999) (Bristol, Great Britan: IOP Publishing)

\bibitem{Backer:1982}D.~C.~Backer, S.~R.~Kulkarni, C.~Heiles {et~al.}, Nature {\bf 300}, 615 (1982).

\bibitem{Cusmano:2004NPPS}G.~Cusumano, W.~Hermsen, M.~Kramer, L.~Kuiper, O.~Lohmer, T.~Mineo, L.~Nicastro, and
B.~W.~Stappers, Nucl. Phys. Proc. Suppl. {\bf 132}, 596 (2004).

\bibitem{Hessels:2006ze}J.~W.~T.~Hessels, S.~M.~Ransom, I.~H.~Stairs, P.~C.~C.~Freire,
V.~M.~Kaspi and F.~Camilo, Science {\bf 311}, 1901 (2006).

\bibitem{Manchester:2005AJ}R. N. Manchester, G. B. Hobbs, A. Teoh, and M.
Hobbs, Astron. J. {\bf 129}, 1993 (2005).

\bibitem{Kaaret:2006gr}P.~Kaaret {\it et al.}, Astrophys. J. {\bf 657}, L97 (2007).

\bibitem{Stergioulas:1994ea}N.~Stergioulas, J.~L.~Friedman, Astrophys. J. {\bf 444}, 306 (1995).

\bibitem{Horowitz:2008vf}
  C.~J.~Horowitz and D.~K.~Berry,
  Phys.\ Rev.\  C {\bf 78}, 035806 (2008).

\bibitem{Laarakkers:1999ApJ}W.~Laarakkers and E.~Poisson, Astrophys. J. {\bf
512}, 282 (1999).
\end{thebibliography}
\end{document}